\documentstyle[aps,pra,preprint]{revtex}

\begin{document}

\title{ Quantum cobwebs: Universal entangling of quantum states}
\author{Arun Kumar Pati$^{(1)(2)}$ \thanks{Email:akpati@iopb.res.in}}
\address{Institute of Physics, Bhubaneswar-751005, Orisaa, India.\\
$^{(1)}$ Center for Philosophy and Foundation of Science, New Delhi, India.\\
$^{(2)}$ School of Informatics, University of Wales, Bangor, LL57 1U
T, UK.}


\maketitle

\begin{abstract}
Entangling an unknown qubit with one type of reference state is generally 
impossible. However, entangling an unknown qubit with two 
types of reference states is possible. To achieve this, we introduce a new 
class of states called {\em zero sum amplitude} (ZSA) multipartite, pure 
entangled states for qubits and study their salient features. Using shared-ZSA 
state, local operation and classical communication we give a protocol for 
creating multipartite entangled states of an unknown quantum state with 
two types of reference states at remote places. This provides a way of 
encoding an unknown pure qubit state into a multiqubit entangled state. 
We quantify the amount of classical and quantum resources required to create 
universal entangled states. This is possibly a strongest form of quantum bit 
hiding with multiparties.
\end{abstract}

\date{\today}
\maketitle
\def\ra{\rangle}
\def\la{\langle}
\def\ver{\arrowvert}

\vskip .5cm


\section{Introduction}
In recent years we have learnt about what we can do and what we cannot do with
the largely inaccessible information content of an unknown quantum state 
\cite{cf}. On the one hand linearity and unitarity of quantum theory are 
guiding principles and on the other hand they put several fundamental 
limitations on quantum information. Some of these limitations are no-cloning
\cite{wz,dd}, no-deleting against a copy \cite{pb}, and no-complementing 
\cite{bhw}. Though exact 
operations are not allowed, these impossible operations can be made possible
with a fidelity at least equal to that of the state estimation fidelity 
\cite{mp}. Processing of the vast amount of information contained in an 
unknown quantum state without destroying the coherence is an important task, 
in general.

Another key feature of the quantum world is the entangled nature of quantum 
states. Though a complete understanding of quantum entanglement is still 
lacking, many important developments have taken place in recent years in
qualifying and quantifying multiparticle quantum entanglement \cite{dag}.
Quantum entanglement is generally regarded as a very useful resource
for quantum information processing \cite{bbps}. 
In a striking discovery it was shown that without violating no-cloning 
principle an {\em unknown} quantum state can be teleported \cite{bbcjpw} with 
unit fidelity from one place to another using a quantum entangled channel
and sending two bits of classical information. This has been also demonstrated
experimentally \cite{db,dbe,af}. Quantum entanglement can be used for dense 
coding \cite{bw}, entanglement swapping \cite{marek,bose}, remote state 
preparation of special qubits \cite{akp}, study of communication cost 
\cite{hlo} and remote state preparation of arbitrary quantum states 
\cite{betal}, 
teleportation of a unitary operator (quantum remote control) \cite{huel}, 
telecloning \cite{murao}, remote information concentration \cite{mv} 
and many more important tasks.  In addition, it is a hope that entanglement 
will play a key role in quantum computation in giving its extra power 
compared to classical computers \cite{mike}.

Here, we find yet another startling application of quantum entanglement.
Imagine that we need to distribute an unknown qubit to more than one party.
If we could distribute to many parties without entangling (i.e. in product 
states) then that would violate no-cloning principle. This means that the 
distributed state of a qubit with $N$ parties  must be in an entangled state. 
But creating an universal entangled state of an unknown state was shown 
to be impossible \cite{bh}, i.e., starting with a state $\ver \psi \ra_1 
\ver 0 \ra_2 \cdots \ver 0 \ra_N$ we cannot create a symmetric universal 
entangled state $\ver \psi \ra_1 \ver 0 \ra_2 \cdots  \ver 0 \ra_N + 
\ver  0 \ra_1 \ver \psi \ra_2  \cdots \ver 0 \ra_N
+ \cdots  \ver 0 \ra_1 \ver 0 \ra_2  \cdots \ver \psi \ra_N$. 
However, the surprising fact is that if we drop symmetric requirement then 
it is possible to create {\em two types of multiparticle entangled state of 
an unknown state}. Precisely, we are looking for a protocol where the 
resulting state will be either $a_1 \ver \psi \ra_1 \ver 0 \ra_2 \cdots  
\ver 0 \ra_N + a_2 \ver  0 \ra_1 \ver \psi \ra_2 \ver 0 \ra_3 \cdots \ver 0 
\ra_N + \cdots  a_N \ver 0 \ra_1 \ver 0 \ra_2  \cdots \ver \psi \ra_N$
or $a_1 \ver \psi \ra_1 \ver 1 \ra_2 \cdots  \ver 1 \ra_N + a_2 \ver  1 \ra_1 
\ver \psi \ra_2 \ver 1 \ra_3 \cdots \ver 1 \ra_N
+ \cdots  a_N \ver 1 \ra_1 \ver 1 \ra_2  \cdots \ver \psi \ra_N$, where an 
unknown state $\ver \psi \ra$ is entangled with reference states $\ver 0 
\ra$ and $\ver 1 \ra$. 
The protocol we present here differs from the original aim of the universal 
entangler as propsed in \cite{bh} because in our scheme 
one can end up with two different type of universal entangled states.
Further, Buzek and Hillery \cite{bh} have studied approximate methods 
to generate universal enatngled states whereas we propose how to 
create exact universal entangled states. This is an important step in the 
sense that our protocol can produce universal entangled states 
(which people have previously thought to be impossible) and which works 
with unit probability of success. It is a hope that exact universal 
entangled states will play important roles for storage of quantum 
information against environmental decoherence \cite{bbdejm}. 

In this paper we introduce a class of entangled states called {\em zero sum 
amplitude} (ZSA) entangled states which may have merit on their own. 
We present a protocol where upon using 
a special class of ZSA multipartite (say $N$, where $N$ is the number of 
parties) 
shared entangled states, local operations and classical communication (LOCC),
one can create two  types of shared-entangled state of an unknown quantum 
state with $(N-1)$ qubits at remote places. The information about an 
unknown state is distributed with all the $N$ parties concerned {\em in
a non-local way}. 
Thus, remote shared-entangling of an unknown state with multiparties is a
very secure way to preserve the information about an unknown state (as long 
as $N$ parties can maintain their quantum correlation). The present scheme 
could have some potential application in multiparty quantum cryptographic 
protocols which will be reported elsewhere \cite{arun}.

The organisation of the present paper is as follows.
In section 2, we introduce the zero sum amplitude entangled state and 
discuss its salient features. In section 3, we give our protocol for 
creating universal entangled states using a tripartite entangled state 
and quantify
the resource needed to do the task. Further, in section 4 we generalise 
it to $N$-partite entangled states and quantify the amount of bipratite 
entanglement and classical communication needed to create $(N-1)$-partite 
quantum cobweb.  We also explain why a classical correlated channel cannot
be used to create universal entangled states. Our example of $N$-partite 
zero sum amplitude entangled state (amplitudes being $N$th root of unity) 
shows that the amount of bipartite splitting entanglement goes as $1/N$ in the 
large $N$ limit and the conclusion follows in section 5.

\section{ Zero sum amplitude entangled states for multiqubits}

For the sake of generality, we introduce an arbitrary pure $N$-qubit zero 
sum amplitude (ZSA) entangled state $\ver \Phi \ra_{12 \ldots N} \in 
{\cal H}^2 \otimes \cdots \otimes {\cal H}^2$ ($N$ times) given by
\begin{equation}
\ver \Phi \ra_{12 \ldots N} =  \sum_{i=1}^{2^N} c_i \ver i \ra_{12 \ldots N},
\end{equation}
where $\{\ver i \ra\}$ is an orthonormal basis for $2^N$-dimensional
Hilbert space, $\sum_{i=1}^{2^N} c_i =0$ (i.e., all the complex amplitudes
sum to zero) and $\sum_{i=1}^{2^N} \ver c_i \ver^2 =1$ (i.e., the state is
normalised to unity). The state space of a quantum system is the complex
projective Hilbert space ${\cal P} = {\cal H}/U(1)$ which can be defined
as a set of rays of the Hilbert space under the projection map $\Pi: {\cal H} 
\rightarrow {\cal P}$. The complex projective Hilbert space has one dimension 
less, i.e., ${\rm dim}{\cal P}= {\rm dim}{\cal H} -1$. For a general ZSA 
state the dimension  of the state space (viewed as a real manifol) is 
$D = (2^{N+1} -3)$ (i.e. a  $D$-dimensional real space) and requires 
$D$ real parameters to sepcify the point on the quantum state space.

In the rest of the paper, we will consider a special class of ZSA states 
where the number of complex amplitudes is equal to the number of parties 
(and since each party posseses one qubit it is also equal to the total 
number of qubits) involved and $N$ orthonormal states contain all zeros 
except at a single entry which contains a one. 
For example, a $N$-partitite ZSA state is given by
\begin{eqnarray}
\ver \Psi \ra_{123 \ldots N} & = & c_1 \ver 100 \ldots 0 \ra_{123 \dots N}
+ c_2 \ver 010 
\ldots 0 \ra_{123 \dots N} + \nonumber \\ 
&&  \cdots  +  c_N \ver 00 \ldots 1 \ra_{123 \dots N}
= \sum_{k=1}^N c_k \ver x_k \ra_{123 \ldots N},
\end{eqnarray}
where $\ver x_k \ra$ ($k=1,2, \ldots N$) is a $N$-bit string containing all 
$0$'s except that $k$th party contains $1$ and the amplitudes obey ZSA 
condition $\sum_k c_k =0$ and the normalisation  $\sum_k \ver c_k \ver^2 
=1$. These class of sates can be completely specified by $(2N-3)$ real 
parameters.

To appreciate the remarkable features of these class of states we first 
discuss the case of two parties. When the number of parties is two, the 
ZSA state is given by
\begin{equation}
\ver \Psi \ra_{12} =  c_1 \ver 10 \ra_{12} + c_2 \ver 01 \ra_{12}.
\end{equation}
The ZSA and normalisation conditions guarantee that the above state is nothing 
but an EPR singlet state $\ver \Psi^- \ra_{12} =  \frac{1}{\sqrt 2}( \ver 10
\ra_{12} - \ver 01 \ra_{12})$, which is just one member of the Bell-sates. 
This state is known to be locally equivalent to other Bell-states and can
be used for succesfull quantum teleportation of an unknown qubit \cite{cb}. 
However throughout the paper whenever we mention multiparticle state  
we will consider three or more qubits, i.e., $N \ge 3$.

Let us introduce a class of tripartite zero sum amplitude normailsed 
entangled state of qubits  $\ver \Psi \ra_{123}  \in {\cal H}^2 \otimes
{\cal H}^2 \otimes {\cal H}^2$ given by
\begin{equation}
\ver \Psi \ra_{123} =  c_1 \ver 100 \ra_{123} + c_2 \ver 010 \ra_{123} +
c_3 \ver 001 \ra_{123},
\end{equation}
where $c_i$'s $(i=1,2,3)$ are non-zero amplitudes of the basis states 
where the $i$th qubit is in the state $\ver 1\ra$. These amplitudes 
obey $\sum_{i=1}^{3} c_i =0$ and $\sum_{i=1}^{3} \ver c_i \ver^2
=1$. Interestingly, states of the type (4), but without ZSA condition, have
shown up in a variety of places in the literature \cite{ckw,cb,bc,dvc}. 
In particular, Coffman {\it et al} \cite{ckw} have shown that these states 
minimise the residual three-tangle. They also show up in the work of 
Brun and Cohen \cite{cb,bc} on GHZ distilation and has been named as 
`triple' states.  Remarkably, D{\"u}r {\it et al} \cite{dvc} have shown 
that any tripartite entangled state can be converted either to the GHZ 
class  or W class of states by LOCC in a probabilistic manner.  
The only difference between the class of tripartite entangled states 
introduced here and the ones studied in \cite{ckw,bc,cb,dvc} is that 
here the amplitudes obey the ZSA condition. We conjecture that the ZSA 
states are not locally equivalent to $\ver GHZ \ra_{123 \ldots N} = 
1/\sqrt 2( \ver 000\ldots 0 \ra_{123\ldots N} +  \ver 111\ldots 1 
\ra_{123\ldots N})$ and $\ver W \ra_{123\ldots N} = 1/\sqrt N \sum_k 
\ver x_k \ra_{123 \ldots N}$ states, except for the trivial case $N=2$. 
Following the discovery of D{\"u}r {\it et al} \cite{dvc} we can say 
that ZSA state may be related to W-states under stochastic LOCC. 
Also one can give a simple proof that the ZSA state 
is inequivalent to $\ver W \ra_{123 \ldots N}$ and $\ver GHZ \ra_{123 
\ldots N}$ states under local unitary operations \cite{note}. 

Note that this tripartite entangled state is not a maximally entangled 
state. The ZSA states are not maximally fragile \cite{gisin}, i.e., 
measurement of any one of the subsystems does not necessarily destroy 
the entanglement between remaining qubits.  For example, if we project 
the first qubit onto computational basis $\ver 0 \ra$, the state of the 
particles $2$ and $3$ is $c_2 \ver 10 \ra_{23} + c_3 \ver 01 \ra_{23}$, 
which is a non-maximally entangled state. But projection onto a basis 
$\ver 1 \ra$ gives a disentangled state. This property 
holds with respect to all other qubits. The one particle reduced density
matrix $\rho_k \in {\cal H}^2$ for any one of the three particles is not 
completely random but a pseudo-pure state given by
\begin{equation}
\rho_k =  \ver c_k \ver^2  I + (1 - 2\ver c_k \ver^2 )\ver 0 \ra  \la 0 \ver , k=1,2,3. 
\end{equation}

Further, if we trace out Alice's qubit, the two-qubit state at Bob and 
Charlie's place is a {\em mixed entangled} state  given by
\begin{eqnarray}
&& \rho_{23} =  \ver c_1 \ver^2 \ver 0 \ra  \la 0 \ver \otimes \ver 0 \ra  
\la 0 \ver  + \ver c_2 \ver^2 \ver 1 \ra  \la 1 \ver \otimes \ver 0 \ra  
\la 0 \ver  + \ver c_3 \ver^2 \ver 0 \ra  \la 0 \ver \otimes  \nonumber \\
&& \ver 1 \ra  \la 1 \ver  +
c_2 c_3^*  \ver 1 \ra  \la 0 \ver \otimes \ver 0 \ra  \la 1 \ver  +
c_2^*c_3  \ver 0 \ra  \la 1 \ver \otimes \ver 1 \ra  \la 0 \ver .
\end{eqnarray}
That (6) is inseparable can be seen by 
applying Peres-Horodecki criterion \cite{peres,horo} which is a necessary 
and sufficient one in ${\cal H}^2 \otimes {\cal H}^2$.  This says that if
a density matrix $\rho$ is separable then the partial transpose has only
nonnegative eigenvalues. If $T$ is a transposition on space of bounded 
operators  ${\cal B}({\cal H})$, then the partial transpose $PT$ (with 
respect to second subsystem) on ${\cal B}({\cal H}) \otimes {\cal B}({\cal H})
$ is defined as $\rho_{m\mu, n\nu}^{\rm PT}= \rho_{m\nu,n\mu}$. Thus, the
partial transpose of two-qubit density matrix (6) is given by
\begin{eqnarray}
\rho_{23}^{\rm PT} =
\left ( \begin{array}{rrrr}
\ver c_{1}\ver^{2} & 0 & 0 & c_{2}^{*}c_{3} \\ 
0 & \ver c_{3} \ver^{2} & 0 & 0  \\
0 & 0 & \ver c_{2} \ver^{2} & 0  \\
c_{2} c_{3}^{*} & 0 & 0 & 0  
\end{array} \right).
\end{eqnarray}

The eigenvalues of $\rho_{23}^{PT}$ are
$ \lambda_1= \ver c_2 \ver^2, ~ \lambda_2= \ver c_3 \ver^2 ,~
\lambda_3= \frac{1}{2}(\ver c_1 \ver^2 + \sqrt{\ver c_1 \ver^4 + 4 \ver c_2 \ver^2 \ver c_3
 \ver^2}) ~$ and $ \lambda_4= \frac{1}{2}(\ver c_1 \ver^2 - \sqrt{\ver c_1 \ver^4 + 4 \ver c_2 
\ver^2 \ver c_3 \ver^2})$.
Though first three eigenvalues are nonnegative the last one is not 
(one can check that $\lambda_3 \lambda_4$ is a negative number). Therefore,
the two qubit density matrix $\rho_{23}$ is inseparable. The same is true if
we trace out any other qubit and look at the density matrix of the 
two-qubit system. A particular measure of entanglement for mixed state is
`entanglement of formation' \cite{bdsw,wkw}. The entanglement of formation of 
$\rho_{23}$ can be computed explicitly and it is given by 
\begin{eqnarray}
&& E_{23}= -\frac{1}{2}(1+\sqrt{1-4 \ver c_2 \ver^2 \ver c_3 \ver^2}) 
\log [ \frac{1}{2}(1+\sqrt{1-4 \ver c_2 \ver^2 \ver c_3 \ver^2} )]  \nonumber\\
&& - \frac{1}{2}(1-\sqrt{1-4 \ver c_2 \ver^2 \ver c_3 \ver^2}) 
\log [ \frac{1}{2}(1-\sqrt{1-4 \ver c_2 \ver^2 \ver c_3 \ver^2})] .
\end{eqnarray}
We will see in the section 3. that the mixed state density matrix (6) is
transformed to a pure state with LOCC. Next we come to the main result of
our paper.

\section{Universal entangling of unknown qubit with two parties}

In what follows we give our protocol for creating two types of 
universal entangled states of an unknown state 
at two remote locations. Suppose Alice, Bob and Charlie at remote locations 
share an entangled state (4) and have access to particles $1,2$ and $3$, 
respectively. An unknown qubit is given to Alice in the form
\begin{equation}
\ver \psi \ra_a = \alpha \ver 0 \ra_a + \beta \ver 1 \ra_a,
\end{equation}
where $\alpha = \cos {\theta \over 2}$ and $\beta = \sin {\theta \over 2} 
\exp (i \phi)$. We show that Alice can always {\em create an entangled 
state of any unknown state with a reference state} $\ver 0 \ra$ or $\ver 1 
\ra$ shared between Bob and Charlie by sending two bits of information to 
both of them. The combined state of the input and the tripartite ZSA 
entangled state $\ver \psi \ra_a \otimes \ver \Psi \ra_{123}$ can be 
expressed in terms of Bell-states \cite{sam} of particle $a$ and $1$ as
\begin{eqnarray}
\ver \psi \ra_a \otimes \ver \Psi \ra_{123} &=& \frac{1}{\sqrt 2} 
\biggl[ \ver \Phi^+ \ra_{a1} \otimes
 ( c_2 \alpha \ver 10 \ra_{23} + c_3 \alpha \ver 01 \ra_{23} + 
\nonumber \\ && 
c_1 \beta \ver 00 \ra_{23}  ) + 
\ver \Phi^{-} \ra_{a1} \otimes
( c_2 \alpha \ver 10 \ra_{23}  + c_3 \alpha \ver 01 \ra_{23}  - 
c_1 \beta \ver 00 \ra_{23} ) \nonumber \\
&& + \ver \Psi^{+} \ra_{a1} \otimes
( c_1 \alpha \ver 00 \ra_{23}  +   c_2 \beta \ver 10 \ra_{23}  + 
c_3 \beta \ver 01 \ra_{23} ) +  
\nonumber \\ && 
\ver \Psi^{-} \ra_{a1} \otimes
( c_1 \alpha \ver 00 \ra_{23}  + c_2 \beta \ver 10 \ra_{23}  - 
c_3 \beta \ver 01 \ra_{23} ) \biggr].
\end{eqnarray}

Using the zero sum amplitude property, i.e., $\sum_i c_i=0$, we can rewrite
the combined state as 
\begin{eqnarray}
&& \ver \psi \ra_a \otimes \ver \Psi \ra_{123} = \frac{1}{\sqrt 2} \biggl[ \ver \Phi^+ \ra_{a1} \otimes
 ( c_2 i \sigma_y  \ver \psi \ra_2 \ver 0 \ra_3 + c_3 \ver 0 \ra i\sigma_y \ver \psi \ra_3 ) + \nonumber \\
&& \ver \Phi^{-} \ra_{a1} \otimes
( c_2  \sigma_x  \ver \psi \ra_2 \ver 0 \ra_3 + c_3 \ver 0 \ra  \sigma_x \ver \psi \ra_3 )   - 
 \ver \Psi^{+} \ra_{a1} \otimes ( c_2  \sigma_z  \ver \psi \ra_2 \ver 0 \ra_3   \nonumber \\
&& + c_3 \ver 0 \ra  \sigma_z \ver \psi \ra_3 ) +  \ver \Psi^{-} \ra_{a1} \otimes
( c_2 \ver \psi \ra_2 \ver 0 \ra_3 + c_3 \ver 0 \ra \ver \psi \ra_3 )  \biggr],
\end{eqnarray}
where $\sigma_x, \sigma_y$ and $\sigma_z$ are Pauli matrices.
Now Alice performs a joint measurement on particles $a$ and $1$. Since a 
Bell-basis measurement will give four possible outcomes $\{ \ver \Phi^{\pm} 
\ra, \ver \Psi^{\pm} \ra \}$ she can get two classical bits of information. 
Then, Alice sends two classical bits to Bob and Charlie both, who in turn can 
apply certain local unitary operations to share an entangled state of an
unknown state with reference states such as $\ver 0 \ra$ or $\ver 1 \ra$. 
For example, if the outcome is $\ver \Phi^{+} \ra $ or $\ver \Phi^{-} \ra $,
then after receiving classical information Bob and Charlie will apply
$i \sigma_y \otimes i \sigma_y$ or $ \sigma_x \otimes  \sigma_x$, 
respectively.  They will be sharing an entangled state given by 
\begin{equation}
\ver \psi^{(1)} \ra_{23} =  c_2 \ver \psi \ra_2 \ver 1 \ra_3 + c_3 \ver 1 \ra \ver \psi \ra_3,
\end{equation}
where $ \ver \psi^{(1)} \ra_{23}$ is an {\em universal entangled state}
of an unknown qubit with a reference state $\ver 1 \ra$. This is universal
because the protocol works perfectly for any input qubit $\ver \psi \ra$.
If the result is $\ver \Psi^{+} \ra $ or $\ver \Psi^{-} \ra $ then, after
receiving classical communication Bob and Charlie will apply $ \sigma_z 
\otimes  \sigma_z$ or $ I \otimes I$, respectively.  In this case they
will be sharing an entangled state given by 
\begin{equation}
\ver \psi^{(0)} \ra_{23} =  c_2 \ver \psi \ra_2 \ver 0 \ra_3 + c_3 \ver 0 \ra \ver \psi \ra_3,
\end{equation}
where the states $\ver \psi ^{(0)} \ra_{23}$ is an {\em universal entangled 
state} of an unknown state with a reference state $\ver 0 \ra$. For
successful creation of universal entangled states 
$\ver \psi^{(0)} \ra_{23}$ or 
$\ver \psi^{(1)} \ra_{23}$ two classical bits are needed from Alice. 
Note that the states in equations (12) and (13) are not normalised.
The normalisation constant for (12) is $ N(\beta) = 1/ \sqrt{ \ver c_2 \ver^2
 + \ver c_3 \ver^3 + 2 \ver \beta \ver^2 {\rm Re} (c_2^*c_3)}$ and for (13) 
is $N(\alpha)$, where $N(\alpha)$ can be obtained from 
$N(\beta)$ by replacing $\beta$ with $\alpha$.

An interesting observation is that if the state $\ver \psi \ra$ is in a known 
state such as $\ver 0 \ra$ or $\ver 1 \ra$, then one may use this scheme
for quantum cryptographic purposes. For example, if $\ver \psi \ra = \ver 0 
\ra$, then $\ver \psi^{(0)} \ra_{23}$ is not an entangled but $\ver 
\psi^{(1)} \ra_{23}$ is. Similarly, if $\ver \psi \ra =  \ver 1 \ra$, then
$\ver \psi^{(1)} \ra_{23}$ is not entangled but $\ver \psi^{(0)} \ra_{23}$ 
is. This may provide a way to generate a coded message (detailed discussions 
are beyond the scope of the present paper and the results will be reported 
elsewhere \cite{arun}).    

\section{Creating quantum cobwebs}

The states that we have created by this protocol are very special.
One can check that there is no local unitary operation ${\cal H}_2 \otimes 
{\cal H}_3$ that can disentangle the unknown state perfectly. Even if both
parties come together and perform joint unitary and measurement operations
they cannot disentangle the qubit perfectly.  Since a general 
quantum operation is a positive, linear, trace preserving map that has a 
unitary representation involving ancilla, let us assume that there is
a unitary operator that disentangles any arbitary 
qubit perfectly. The action of the unitary operator on universal entangled 
state of $\ver \psi \ra$ and $\ver {\bar \psi} \ra= \alpha \ver 1 \ra - 
\beta^* \ver 0 \ra $ (with $\la \psi \ver {\bar \psi} \ra =0$) will be given by
\begin{eqnarray}
&&  N(\alpha) (c_2 \ver \psi \ra_2 \ver 0 \ra_3 + c_3 \ver 0 \ra \ver \psi \ra_3) \ver A \ra \rightarrow \ver 0 \ra \ver \psi \ra \ver A '\ra, \nonumber \\
&&  N(\beta) (c_2 \ver {\bar \psi} \ra_2 \ver 0 \ra_3 + c_3 \ver 0 \ra \ver {\bar \psi} \ra_3) \ver A \ra 
\rightarrow \ver 0 \ra \ver {\bar \psi} \ra \ver A ''\ra,
\end{eqnarray}
where $\ver A \ra$ is the initial and $\ver A '\ra, \ver A ''\ra$ are the 
final states of the ancilla, $N(\alpha)$ and  $N(\beta)$ are normailsation 
constants for entangled states of $\ver  \psi^{(0)} \ra$ and
$\ver {\bar \psi}^{(0)} \ra$, respectively. Taking the inner product 
we have $2 N(\alpha) N(\beta) \alpha \beta^* {\rm Re}(c_2c_3^*)=0$ and this
can never be satisfied for any non-zero values of $c_2, c_3, \alpha$ and
$\beta$. Therefore, we cannot disentangle the state even by joint action and
irreversible operation. Thus, the unknown state (containing some secret 
information) can remain simultaneously with two parties in a non-local 
manner. The element of surprise is in the fact that this is seems to be an 
irreversible conversion process (somewhat analogous to other irreversible 
distillation process.) 
This class of states we call {\em quantum cobwebs} because once they are
created the unknown state is trapped in side the multiparticle entangled
state.
Though this feature may look 
undesirable to some readers it is indeed very useful in quantum 
cryptographic schemes. Often, new quantum information processing protocols are 
double-edged swords. If there is a negative aspect of a protocol there is a 
great positive aspect as well. However, here we would like to leave it
as an open question  whether these universal entangled states can be 
disentangled perfectly using entanglement assisted local operation and 
classical communication.

As mentioned in the introduction, recently it was shown \cite{bh} that 
there is no unitary operator which can create a perfect symmetric universal 
entangled state that will take $\ver \psi \ra_1 \ver 0 \ra_2 
\rightarrow  \ver \psi \ra_1 \ver 0 \ra_2 +  \ver 0 \ra_1 \ver \psi \ra_2$.
Similarly, the reverse operation, i.e., a perfect disentangler is also not 
possible \cite{bh1}. But in our protocol we have circumvented this limitation 
and achieved two types of arbitrary {\em universal perfect entanglers} with 
unit probability using shared entanglement, local operations and classical 
communications (LOCC). However, our universal entangled states are not 
permutationaly invariant. Of course, our scheme may not be the only way to 
generate universal entangled states. It could be possible to consider 
a unitary operation on an 
unknown state along with ancillas and one may be able to create two types of 
universal entangled states with a postselection of 
measurement result. But this needs further investigation.

It may be worth mentioning that if Bob and Charlie perform non-local 
unitary operation and measurement, then one of them can recover the state 
with unit fidelity in a {\em probabilistic manner}. For example, to 
disentangle $\ver \psi^{(0)} \ra_{23}$ Bob and Charlie can come together and 
perform a CNOT operation followed by a measurement of particle $2$ in the 
basis $\{ \ver + \ra= \frac{1}{\sqrt 2}(\ver 0 \ra + \ver 1 \ra), \ver - 
\ra =  \frac{1}{\sqrt 2}(\ver 0 \ra - \ver 1 \ra) \} $. When Bob gets $\ver 
+ \ra$, Charlie's qubit is in the state $\ver \psi \ra$ 
and when Bob gets $\ver - \ra$ Charlies qubit is not in the state $\ver \psi 
\ra$, (i.e. there is an error in getting $\ver \psi \ra$) so they can 
discard this. The probability of success is $P= [ \ver c_2 \ver^2 + \ver c_3 
\ver^3 + 2 {\rm Re}(c_2^*c_3)]/ 2[ \ver c_2 \ver^2 + \ver c_3 \ver^3 + 2 
\alpha^2 {\rm Re}(c_2^*c_3)]$ which is greater than half (i.e. better than 
a random guess).  

\section{Resource for universal entangled states}

Next, we quantify the amount of nonlocal quantum resource needed to create 
a remote shared-entangled state.  In general quantifying amount of 
entanglement in multiparticle system is still a difficult problem \cite{vw}. 
Moreover, for more than two parties there is no unique measure of
quantum entanglement \cite{vidal}. However, our purpose is not to provide a 
measure of entanglement for ZSA sates. We simply observe the following points. 
The tripartite system can be partioned in three different ways, 
i.e $A$ vs $BC$, $B$ vs $AC$ and $C$ vs  $AB$, there are three different 
ways of calculating the bipartite entanglement. Since we are interested to 
know the quantum resources between $A$ versus $BC$ (as we are creating 
universal entanglers for $BC$), we look at the amount of bipartite 
entanglement with respect to splitting of particles $A$ vs $BC$. This is 
given by the von Neumann entropy of the reduced density matrix 
$\rho_1$ \cite{pr}
\begin{eqnarray}
&& E(A~{\rm vs}~ BC) = E(\rho_1)  =  - {\rm tr}(\rho_1 \log \rho_1) =  \nonumber \\
&& -  (1-\ver c_1 \ver^2) \log(1-\ver c_1 \ver^2)  -
\ver c_1 \ver^2 \log \ver c_1 \ver^2. 
\end{eqnarray}
Though, the use of von Neumann entropy as a measure of entanglement for 
bipartite system is justified when asymptotically large number of copies are
involved \cite{gv}, we use this for its simplicity. Therefore, we can roughly 
say that using $E(\rho_1)$ amount of entanglement and communication of 
two classical bits to Bob and Charlie one can create two types of quantum 
universal entanglers for an unknown state. Thus a mixed entangled state 
(6) is converted 
to  a pure universal entangled state after receiving classical communication 
from Alice. (Recall a similar situation in quantum teleportation,
where a completely random mixture is converted to a pure unknown state). 
This is very interesting process where a mixed 
entangled state shared between two parties Bob and Charlie is purified to a 
pure entangled state using LOCC along with assisted LOCC from a third party 
Alice.

We can also argue that no classical correlated state (CCS) can create an 
universal entangled state of an unknown state. If we could create an 
univesral entangled state using CCS via local operation and classical 
communication then we could create some amount of entanglement between Bob 
and Charlie. But we know that via LOCC one cannot create any entanglement
\cite{vidal}, hence CCS cannot create any universal entangled states. 

We can actually quantify the amount of entanglement present in a bipartite 
universal entangled state (quantum cobweb). When the universal entangled 
state is of the type (13), then the reduced density matrix of qubit $2$ 
at Bob's place is 
\begin{eqnarray}
&& \rho_{2} =  N(\alpha)^2 [\ver c_2 \ver^2 \ver \psi \ra  \la \psi \ver  +
\ver c_3 \ver^2 \ver 0 \ra  \la 0 \ver +
c_2 c_3^* \alpha \ver \psi \ra  \la 0 \ver  + 
c_2^*c_3  \alpha \ver 0 \ra  \la \psi \ver ].
\end{eqnarray}

In the bipartite case the Schmidt decomposition theorem \cite{ep} guarantees 
that the eigenvalues of the reduced density matrices of B and C will be 
identical. They are given by $\eta_{\pm} = \frac{1}{2}(1 \pm 
\sqrt{1 - 4 \epsilon})$, where $\epsilon = 4 N(\alpha)^4 \ver \beta 
\ver^4 \ver \ver c_2 \ver^2 \ver \ver c_3 \ver^2 $.
Therefore, the amount of entanglement will be 
\begin{eqnarray}
E(\ver \psi^{(0)} \ra_{23}) =  - {\rm tr}(\rho_2 \log \rho_2) = 
 - {\rm tr}(\rho_3 \log \rho_3) = -  \eta_+ \log \eta_+  - \eta_- \log \eta_-. 
\end{eqnarray}

As an example, if the amplitudes are cube roots of unity, then the zero sum amplitude entangled state is of the form 
\begin{equation}
\ver \Psi \ra_{123} =  \frac{1}{\sqrt 3} ( \ver 100 \ra_{123} + e^{2\pi i/3} \ver 010 \ra_{123} + e^{-2\pi i/3} \ver 001 \ra_{123}).
\end{equation}
The reduced density matrices for each of the subsystem are same and also 
has equal spectrum. It is given by $\rho_{1} = \rho_{2} = \rho_{3} ={\rm diag}
( 2/3, 1/3 )$. Therefore, the amount of bipartite entanglement between any 
partioning is $E(\rho_1)  =  E(\rho_2) = E(\rho_3) = 1-(5-3 \log 3)/3 = .9  
{\rm ebits}$. Thus, with a use of $.9$ ebits of entanglement and two cbits 
of communication one can create, for example, a universal entangled state 
of the form
\begin{equation}
\ver \Psi^{(0)} \ra_{23} =  \frac{1}{\sqrt 3} ( e^{2\pi i/3} \ver  \psi \ra_2 \ver 0 \ra_3 + 
e^{-2\pi i/3} \ver 0 \ra_2 \ver \psi \ra_3). 
\end{equation}

\section{Universal entangled state for multiparties}

We can generalise the universal quantum entangler for $(N-1)$ parties 
where an unknown qubit can be entangled with a reference state and shared 
with $(N-1)$ parties. Let there be $N$ parties in a network of $N$ nodes 
each having access to a single qubit. They share $N$-partite zero sum 
amplitude entangled state $\ver \Psi \ra_{123 \ldots N}  
\in {{\cal H}^2}^{\otimes N}$  given by (2).

Now, we describe how Alice can create an $(N-1)$-partite entangled state of 
any unknown state with a reference state $\ver 0 \ra$ or $\ver 1 \ra$ shared 
between Bob, Charlie....and Nancy by sending two bits of information to the 
concerned parties. The combined state of the unknown 
qubit and $N$-partite entangled state $\ver \psi \ra_a \otimes \ver \Psi 
\ra_{123 \dots N}$ can be expressed in terms of Bell-states of particle $a$ 
and $1$ as (again using the zero sum amplitude property) 
\begin{eqnarray}
&& \ver \psi \ra_a \otimes \ver \Psi \ra_{123 \ldots N} = \frac{1}{\sqrt 2} \biggl[ \ver \Phi^+ \ra_{a1} \otimes
\sum_{k=2}^N c_k  \ver (i \sigma_y )  \psi_{(k)}^{(0)} \ra_{23 \ldots N}  +  \nonumber \\
&& \ver \Phi^{-} \ra_{a1} \otimes
\sum_{k=2}^N c_k  \ver  (\sigma_x )  \psi_{(k)}^{(0)}  \ra_{23 \ldots N}  
 - \ver \Psi^{+} \ra_{a1} \otimes 
 \sum_{k=2}^N c_k  \ver  (\sigma_ z) \psi_{(k)}^{(0)}  \ra_{23 \ldots N}  +  \nonumber \\
&& \ver \Psi^{-} \ra_{a1} \otimes
\sum_{k=2}^N c_k  \ver  \psi_{(k)}^{(0)}  \ra_{23 \ldots N}
\biggr],
\end{eqnarray}
where $\ver \psi_{(k)}^{(0)}  \ra_{23 \ldots N} = 
\ver 0 \ra_2 \ver 0 \ra_3 \cdots \ver \psi \ra_k \cdots \ver 0 \ra_N$ 
is a $(N-1)$ qubit strings containing all qubits in the state $\ver 0 \ra$ 
except that the $k$th party contains the unknown state $\ver \psi \ra$.
Alice performs a joint Bell-state measurement on particles $a$ and $1$. 
If the outcome is  $\ver \Phi^{+} \ra $ or $\ver \Phi^{-} \ra $ then after 
sending classical communication to the concerned  $(N-1)$ parties, they will 
apply $i \sigma_y \otimes \cdots \otimes i \sigma_y$ or $ \sigma_x \otimes 
\cdots \otimes  \sigma_x$, respectively. They will end up 
sharing an entangled state given by 
\begin{equation}
\ver \psi^{(1)} \ra_{23 \ldots N} =  \sum_{k=2}^N c_k \ver \psi_{(k)}^{(1)}  \ra_{23 \ldots N},    
\end{equation}
where $\ver \psi_{(k)}^{(1)}  \ra_{23 \ldots N}=
\ver 1 \ra_2 \ver 1 \ra_3 \cdots \ver \psi \ra_k \cdots \ver 1 \ra_N$ 
is a $(N-1)$ qubit strings that contains all 
qubits in the state $\ver 1 \ra$ except that the $k$th party contains the 
unknown state $\ver \psi \ra$. If the outcome is  $\ver \Psi^{+} \ra $ or 
$\ver \Psi^{-} \ra $ then after receiving classical communication, $(N-1)$ 
parties will apply $ \sigma_z \otimes \cdots \otimes 
\sigma_z$ or $ I \otimes \cdots \otimes I$ (do nothing), respectively.  
Thus, they will end up sharing an entangled state given by 
\begin{equation}
\ver \psi^{(0)} \ra_{23 \ldots N} =  \sum_{k=2}^N c_k \ver \psi_{(k)}^{(0)}  \ra_{23 \ldots N}.
\end{equation}
Thus, with the use of zero sum amplitude entangled state and two classical 
bits one can create universal entangled states of an unknown state with two 
types of reference states that have been shared between $(N-1)$ parties at 
remote locations. Thus,  $\ver \psi^{(0)} \ra_{23 \ldots N}$ and $\ver 
\psi^{(1)} \ra_{23 \ldots N}$ are $(N-1)$ node quantum cobwebs from 
which an unknown state cannot be disentangled perfectly by local or nonlocal 
unitary operations. The multiparty cobweb states in (21) and (22) are not 
normalised as expected.

The reduced density matrices for single qubits (after tracing out other 
$(N-1)$ qubits from the $N$-partite state (2) ) is given by (5) with 
$k=1,2, \ldots N$. With $N$ parties there are $N(N-1)/2$ possible bipartite 
entanglements but we are interested in entanglement with respect to $N$ 
number of bipartite partioning (where we make partioning of one qubit versus 
all other qubits). The amount of bipartite entanglement with 
respect to splitting between qubit $1$ vs $(N-1)$ is again given by (15). 
Therefore, with the use of $E(\rho_1)$ amount of entanglement and two 
classical bits to $(N-1)$ parties one can create two types of $(N-1)$ 
partite universal entangled states. 

As an example, if the amplitudes are $N$th roots of unity, then the zero sum amplitude entangled state is given by
\begin{eqnarray}
\ver \Psi \ra_{123 \ldots N} = \frac{1}{\sqrt N}\sum_{k=1}^N e^{i 2\pi k/N}  \ver x_k \ra_{123 \ldots N}.
\end{eqnarray}
The reduced density matrix of any of the qubit is identical and is given 
by $\rho_k = {\rm diag}[( 1-1/N), 1/N]$. Therefore, the amount of bipartite 
entanglement is independent of the choice of  $N$ possible bipartite 
partioning. The bipartite entanglement  $E$ with respect to splitting 
between particle $1$ and the rest $(N-1)$ qubits is
\begin{eqnarray}
E = E(\rho_1)  =  -[(1- \frac{1}{N}) \log(1- \frac{1}{N})  +  \frac{1}{N} \log \frac{1}{N} ]
\end{eqnarray}

With the use of $E$ bits of entanglement one can create an universal 
entangled state of $(N-1)$ qubits as
\begin{eqnarray}
\ver \psi^{(0)} \ra_{23 \ldots N} = \frac{1}{\sqrt N}\sum_{k=2}^N e^{i 2\pi k/N}  \ver \psi_{(k)}^{(0)} \ra_{23 \ldots N}.
\end{eqnarray}

If the number of parties $N$ becomes very large $E \rightarrow 1/N$, 
this approaches zero  i.e., the bipartite entanglement for the state (23) 
cannot be unlimitedly distributed between large number of parties. For 
large but finite $N$ we can say that with the use of $O(1/N)$ 
ebits of bipartite entanglement we can prepare a universal entangled state
(25) for 
an unknown state with $O(N)$ parties at remote locations. In a different 
context, it was shown \cite{kbi} that bipartite entanglement distributed 
between $N$ parties goes as $2/N$.

\section{Conclusion}

In this paper, we have introduced a class of zero sum amplitude multipartite 
entangled states and studied their properties. 
Interestingly, when the number of parties is two, 
the ZSA entangled state is exactly an EPR state. We have presented a 
protocol where one can create two types of universal entangled states of 
an unknown state with reference states using shared ZSA 
entangled states and LOCC, which was thought to be an impossible task. 
This class of states 
may be called quantum cobwebs. This surprising feature exploits one 
property that is the zero sum amplitude nature of the original shared 
entangled state between $N$ parties. 
Creating a quantum cobweb could have some strategic 
applications, where some secret information is shared with every body 
but no one can salvage that information. This is very useful for 
cryptographic schemes. It may be remarked that though the original 
quantum teleportation uses maximally bipartite entangled states, one 
can also use three particle and four particle GHZ states for quantum 
teleportation \cite{bour,akp1}. Ours is one example, where these class of 
multipartite states are pure entangled states but {\em are not useful} for 
quantum teleportation. We hope that the nature of ZSA and universal 
entangled states will throw some new light on the nature of quantum 
information and role of entanglement. Some open questions include: 
Is our scheme the most simple scheme for creating cobwebs? Why the ZSA
states are so special and can one create cobwebs with other entangled states?
In future, one can also explore if these 
multipartite ZSA entangled states can be employed for some other quantum 
information processing tasks.

\vskip .5cm

\noindent
{\bf Acknowledgements:} 
I thank S. Bose for very useful discussions and A. Chefles for useful 
remarks. I thank S. R. Jain, Z. Ahmed, H. D. Parab for discussions 
concerning the zero sum nature of complex numbers. 
I also thank T. Brun for useful remarks and bringing some useful references
to my notice.



\begin{thebibliography}{99}
\bibitem{note}
Suppose, there is a local unitary operator $U=U_1 \otimes U_2
\cdots U_N$ such that $U \ver W \ra_{123 \ldots N} = \ver \Psi \ra_{123
\ldots N}$. This implies that the amplitudes obey  $c_l = 1/\sqrt N \sum_k
U_{lk}$ and zero sum amplitude condition implies $\sum_{lk}U_{lk}=0$. First,
consider the case where the unitray operator is uni-local (unitary operator
acting only on one party), i.e., $U=U_1 \otimes I_2 \cdots I_N$. The general
case would be obtained by composing uni-local unitary operators $U_1 \otimes 
I_2 \cdots I_N$ with  $I_1 \otimes U_2 \cdots I_N$
and so on \cite{dvc}. Inserting $U$, we have the equation $\la 1 \ver U_1 \ver 1 \ra + (N -1 ) \la 0 \ver U_1 \ver 0 \ra 
=0$. But this condition can never be satisfied by any local unitary operator. 
By composing uni-local unitary operators one can see that there is no 
arbitrary local unitary operator such that $\ver \Psi \ra$and $\ver W \ra$ 
are locally equivalent.  Similarly, one can show that $\ver \Psi \ra$ and
$\ver GHZ \ra$ states are not related by local unitary operators for $N \ge
3$. If they are, then zero sum amplitude condition implies that $\la 1 
\ver U_1 \ver 0 \ra +  \la 1\ver U_1 \ver 1\ra =0$ and this can never
be satisied. This proves that the ZSA state is not equivalent to $\ver
W \ra$ and $\ver GHZ \ra$ states under local unitary operators.
\end{thebibliography}
\end{document}